hep-lat/9605022  18 May 1996

# Investigation of spontaneous symmetry breaking from a non standard approach

Vicente Azcoiti, Victor Laliena,[a] and Xiang-Qian Luo[b]

[a]Departamento de Física Teórica, Universidad de Zaragoza, 50009 Zaragoza, Spain

[b]HLRZ (Supercomputing Center), Forschungszentrum, D-52425 Jülich, Germany,
and Deutsches Elektronen-Synchrotron DESY, D-22603 Hamburg, Germany

We propose a new method for the study of the chiral properties of the ground state in Quantum Field Theories (QFT′s) which is based on the computation of the probability distribution function (p.d.f.) of the chiral condensate in the chiral limit. We show how despite of the fact that Grassmann variables cannot be simulated on a computer, an analysis of spontaneous symmetry breaking without a symmetry breaking external field, which is standard in the case of spin systems, can also be done in QFT′s with fermion degrees of freedom.

Chiral-symmetry breaking ($\chi$SB) is an important feature of many systems with fermions. In QCD at finite temperature, there is a chiral phase transition from a confining phase, where chiral symmetry is spontaneously broken to the deconfining phase, where chiral symmetry is restored. In non-compact QED, a chiral transition separates the broken phase and Coulomb phase, raising the possibility of the existence of the continuum limit.

Although these issues have been extensively investigated for a long time, there are still many open questions. The main question we would address here is how to get quantitative information on spontaneous chiral-symmetry breaking from numerical simulations.

In the chiral limit, if we use the chiral symmetric action and boundary conditions satisfying the symmetry requirement, the value of any order parameter is always zero, and we don't know in which phase the system is. The usual way is to add to the action a symmetry breaking term so that the fermion condensate is non-zero. An extrapolation to the chiral limit has to be done after the thermodynamical limit. Unfortunately, such an extrapolation in numerical analysis is an extrapolation from data obtained on a finite lattice at small set of fermion masses to the massless limit, in which the operation is ambiguous and therefore the result might be arbitrary. Can we avoid such an ambiguous extrapolation and obtain information directly from the symmetric action?

It is well known for bosons or a spin system [1], one can work directly with a symmetry action. From the data of a simulation, one can construct the probability distribution function (p.d.f.) of the order parameter from which the value of the order parameter can be estimated by finite size scaling [2].

For fermions, this method is not straightforward. The Grassmann variables can not directly be simulated on the computer and Mathews-Salam formula has to be used to integrate out the fermionic degrees of freedom. Due to such an integration, in the chiral limit, the expectation value for any order parameter vanishes identically. Then it is not possible to construct the p.d.f of the order parameter as a histogram. But this doesn't mean we can not obtain information on spontaneous $\chi$SB (S$\chi$SB) from the p.d.f.

First of all, let's introduce some basic concepts in our approach. When a symmetry is spontaneously broken, a set of degenerate pure vacua appears, forming a Gibbs state, i.e., a macroscopic state. The expectation value for any operator $O$ in the Gibbs state can be written as

$$<O> = \sum_\alpha w_\alpha <O>_\alpha, \qquad (1)$$

where $\alpha$ labels the pure vacuum, and $w_\alpha$ is the probability of the pure vacuum $\alpha$ in the Gibbs state. Let us choose the chiral condensate $\bar\psi\psi$ as the order parameter, then the expectation value of the order parameter in the $\alpha$ vacuum is

$$c_\alpha = \frac{1}{V}\sum_x <\bar\psi(x)\psi(x)>_\alpha. \qquad (2)$$

All pure vacua, and only pure vacua, satisfy the cluster properties, which implies that the intensive quantities like the magnetization or chiral condensate do not fluctuate in the pure vacua.

With these basic concepts, we can define the "macroscopic" p.d.f. of the order parameter $c$ by

$$P(c) = \sum_\alpha w_\alpha \delta(c - c_\alpha). \qquad (3)$$

The function $P(c)$ tells us what is the probability that choosing RANDOMLY a vacuum state, we get the value $c$ for the chiral order parameter. If the vacuum state is invariant under chiral transformations, i.e., if it is unique as concerning the chiral symmetry, $P(c)$ will be a single $\delta$ function $\delta(c)$. If there is S$\chi$SB, $P(c)$ will be a complicate distribution.

Let's go further to look at the form of $P(c)$. Since we are interested in the analysis of the chiral symmetry on the lattice, we will use the staggered fermion regularization. In such a case, there is a continuous U(1) chiral symmetry in the action. Then if this continuous residual symmetry is spontaneously broken, we will get a continuous set of equilibrium states labeled by an angle $\alpha \in [0, 2\pi]$. The $v.e.v.$ of the chiral condensate $c_\alpha$ at each $\alpha$ vacuum can be parameterized as $c_0 \cos(\alpha)$, $c_0$ being the value corresponding to the $\alpha - vacuum$ selected when switching-on an external *"magnetic"* field and taking the massless limit afterwards. In another word, $c_0$ is the chiral order parameter in the conventional sense. Notice that no matter in what phase the system is, the integration of the fermionic degrees of freedom (equivalently the integration over the pure vacua) always leads to $<c>=0$. The function $P(c)$ can be computed as

$$P(c) = \frac{1}{2\pi}\int_0^{2\pi} d\alpha \delta(c - c_0\cos(\alpha)). \qquad (4)$$

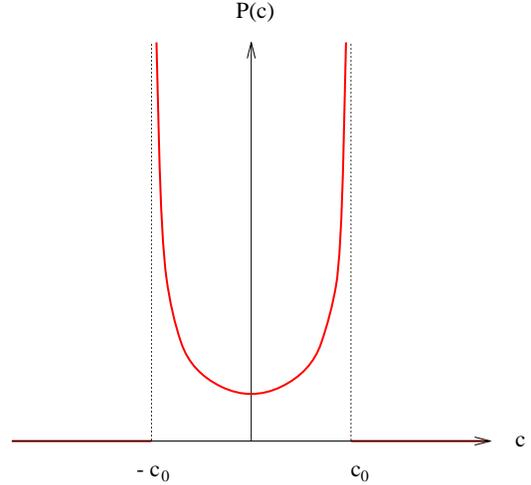

Figure 1. Standard form of $P(c)$ of (4) in the broken phase.

In the symmetric phase $c_0 = 0$, it is obvious that $P(c) = \delta(c)$. In the broken phase, $P(c) = 1/(\pi(c_0^2 - c^2)^{1/2})$ for $-c_0 \leq c \leq c_0$, and $P(c) = 0$ otherwise (see Fig. 1). Its Fourier transformed

$$\tilde{P}(q) = \frac{1}{2\pi}\int_0^{2\pi} d\theta e^{iqc_0\cos\theta} = J_0(qc_0). \qquad (5)$$

Let us emphasis the power of p.d.f.: although $<c>=0$ in all cases, from the shape of p.d.f., we can still determine in which phase the system is and what value of the order parameter $c_0$ would be.

On a finite lattice, let us define p.d.f. as

$$P_V(c) = <\delta(\frac{1}{V}\sum_x \bar\psi(x)\psi(x) - c)> \qquad (6)$$

where the expectation value in (6) is computed in the Gibbs state and the integration measure is that associated to the partition function

$$\mathcal{Z} = \int d\bar\psi d\psi dU e^{-S_G(U) + \bar\psi\Delta\psi}. \qquad (7)$$

$S_G$ in (7) is the pure gauge action and $\Delta$ the fermionic matrix.

Using the cluster properties of the pure vacua, it can be shown [3] that the definition (6) is identical to (3) in the thermodynamical limit, i.e.,

$$lim_{V \to \infty} P_V(c) = P(c). \qquad (8)$$

Expression (6) is not suitable for numerical computation. However its Fourier transformed $\tilde{P}(q)$ as will be shown, can be numerically computed. The fermion matrix $\Delta$ can be decomposed as $\Delta = m + i\Lambda$ where $m$ is the fermion mass and $\Lambda$ a hermitian matrix which depends on the gauge field configuration. The eigenvalues of $\Lambda$ are real and symmetric. Taking into account all these properties of $\Delta$, in the chiral limit the following expression for $\tilde{P}_V(q)$ can be derived

$$\tilde{P}_V(q) = < \prod_j (1 - \frac{q^2}{N^2 \lambda_j^2}) > \qquad (9)$$

where the product in (9) runs over all positive eigenvalues $\lambda_j$ and the mean values are computed with the probability distribution function of the effective gauge theory obtained after integrating out the fermion fields. The function $\tilde{P}_V(q)$ can be computed numerically and then, by inverse Fourier transform we get $P_V(c)$.

Finite size scaling of (9) allow us to get quantitative information on S$\chi$SB. If the symmetry is not broken, in the thermodynamical limit, the right hand side approaches 1. Otherwise, according to (5) and (9),

$$lim_{V \to \infty} < \prod_j (1 - \frac{q^2}{N^2 \lambda_j^2}) > = J_0(qc_0). \qquad (10)$$

This equation also holds for a given background gauge field $U$, from which we obtain for this given configuration the relation between the lattice chiral condensate $c_0(U)$ and $i$th zeroes $a_i$ of $J_0$

$$c_0(i, U) = \frac{a_i}{V \lambda_i}. \qquad (11)$$

According to finite size scaling, if the small eigenvalues go to zero slower than $1/V$, then $c_0(U) =$

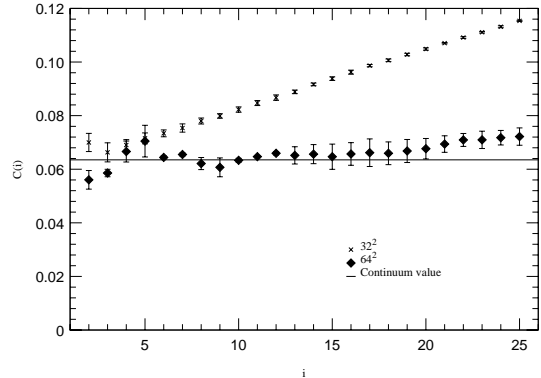

Figure 2. Chiral order parameter in the compact Schwinger model in $32^2$ and $64^2$ lattices at $1/g^2 = 6.344$. The solid line corresponds to its continuum analytical value.

0, i.e., chiral symmetry is restored for this configuration. If they approach zero as $1/V$ then in the thermodynamical limit, then we expect S$\chi$SB. Of course, on finite volume, it is not surprising that $a_i/(V\lambda_i)$ depends on $i$. However in broken phase, a plateau for various $i$ should appear. The extent of the plateau increases with the lattice size. Averaging $c_0(i, U)$ over different configurations, we get the value of the chiral condensate.

These properties are nicely demonstrated in Fig. 2 where we plot the lattice chiral condensate $c_0(i)$ against $i$ for the one-flavor compact Schwinger model at $1/g^2 = 6.344$. The continuous line in this figure stands for the continuum analytical result ($< \bar{\psi}\psi > /e \approx 0.16$) times the bare coupling $g$. Figure 2 tells us that the extent of plateau indeed increases with the lattice size and approaches the exact value in the continuum. Similar behaviors have also been observed for other $g$ values in the weak coupling region. Although the Schwinger model is not physically realistic, it has been widely accepted as a very good laboratory to check new proposals. This is well known for two reasons: the one-flavor massless Schwinger model in the continuum limit is exactly solvable and it shares many interesting properties with other more relevant physical models. Therefore, the results for the Schwinger model strongly

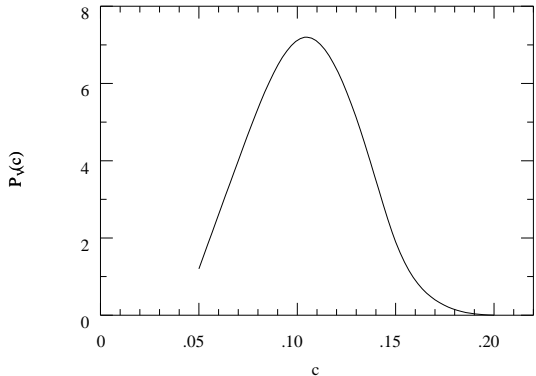

Figure 3. $P_V(c)$ in 4 flavor non-compact QED at $\beta = 0.200$ on $10^4$ lattice.

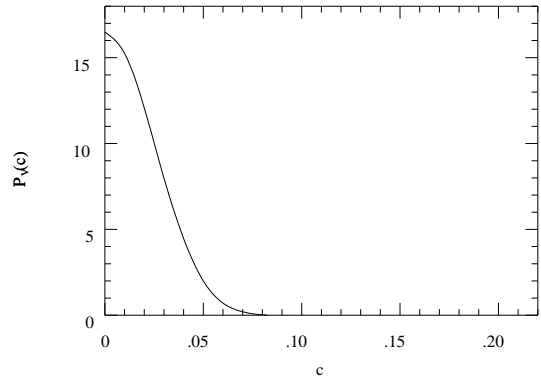

Figure 4. $P_V(c)$ in 4 flavor non-compact QED at $\beta = 0.237$ on $10^4$ lattice.

support the reliability of our method.

Contrary to the asymptotic free theories, the continuum limit of non-compact QED is defined through the chiral phase transition at some finite $\beta$. The issue of triviality (or non-triviality) of the continuum limit is still under debate (for an overview or references see [4,5]). The main difficulty in establishing the nature of the continuum limit is in determination of the critical exponents, since it is very sensitive to the critical coupling $\beta_c$. Recently, the measurements of the susceptibilities [6] suggest $\beta_c \approx 0.202$ for 4 flavors. It is very important to have more independent checks on it. For this purpose, we extend the p.d.f. analysis to this model. Only the results for 4 flavors on $10^4$ lattice will be presented here. Figure 3 shows $P_V(c)$ at $\beta = 0.200$. A peak away from the origin develops (also for $P_V(-c)$ since it is a symmetric function). According to the above arguments, at this $\beta$, the system is in the broken phase. $P_V(c)$ at $\beta = 0.237$ is shown in Fig. 4. From the distribution, we expect that the system is in the symmetric phase at this $\beta$. Therefore figures 3 and 4 give a consistent picture with [6]. Of course, according to (8), computation on different lattices seems necessary.

In conclusion, we have developed a new quantitative approach to spontaneous chiral-symmetry breaking. The main advantage of this method, when compared with standard simulations, is that we can work directly in the chiral limit and therefore no mass extrapolations are needed. From the probability distribution function of the order parameter and finite size scaling, we can obtain not only the information on S$\chi$SB, but also the value of the chiral condensate. We have tested this method in the Schwinger model and applied it to QED. These results should stimulate people working in this field to apply this formalism to more interesting physical systems, like QCD.

VA and VL are partly supported by CICYT (Spain) and XQL is employed by DESY. XQL also thanks the organizers of LATTICE 95 for partial financial support.